\documentclass[twocolumn]{aastex631}
\usepackage[caption=false]{subfig}

\shorttitle{Reading Between the Lines}
\shortauthors{Ostberg et al.}

\begin{document}

\title{Reading Between the Lines: Investigating the Ability of JWST to Identify Discerning Features in exoEarth and exoVenus Transmission Spectra}

\author[0000-0001-7968-0309]{Colby Ostberg}
\affiliation{Department of Earth and Planetary Sciences, University of California, Riverside, CA 92521, USA}
\email{costb001@ucr.edu} 

\author[0000-0002-7084-0529]{Stephen R. Kane}
\affiliation{Department of Earth and Planetary Sciences, University of California, Riverside, CA 92521, USA}

\author[0000-0003-0429-9487]{Andrew P. Lincowski}
\affiliation{Department of Astronomy and Astrobiology Program, University of Washington, Box 351580, Seattle, Washington 98195, USA}
\affiliation{NASA NExSS Virtual Planetary Laboratory, Box 351580, University of Washington, Seattle, Washington 98195, USA}

\author[0000-0002-4297-5506]{Paul A.\ Dalba}
\altaffiliation{Heising-Simons 51 Pegasi b Postdoctoral Fellow}
\affiliation{Department of Astronomy and Astrophysics, University of California, Santa Cruz, CA 95064, USA}
\affiliation{SETI Institute, Carl Sagan Center, 339 Bernardo Ave, Suite 200, Mountain View, CA 94043, USA}


\begin{abstract}

The success of the Transiting Exoplanet Survey Satellite (TESS) mission has led to the discovery of an abundance of Venus Zone (VZ) terrestrial planets that orbit relatively bright host stars. Atmospheric observations of these planets play a crucial role in understanding the evolutionary history of terrestrial planets, past habitable states, and the divergence of Venus and Earth climates. The transmission spectrum of a Venus-like exoplanet can be difficult to distinguish from that of an Earth-like exoplanet however, which could severely limit what can be learned from studying exoVenuses. In this work we further investigate differences in transmission between hypothetical exoEarths and exoVenuses, both with varying amounts of atmospheric carbon dioxide (CO$_2$). The exoEarths and exoVenuses were modelled assuming they orbit TRAPPIST-1 on the runaway greenhouse boundary. We simulated James Webb Space Telescope (JWST) Near-Infrared Spectrograph (NIRSpec) PRISM transit observations of both sets of planets between 0.6--5.2 $\mu$m, and quantified the detectability of major absorption features in their transmission spectra. The exoEarth spectra include several large methane (CH$_4$) features that can be detected in as few as 6 transits. The CH$_4$ feature at 3.4 $\mu$m is the optimal for feature for discerning an exoEarth from an exoVenus since it is easily detectable and does not overlap with CO$_2$ features. The sulfur dioxide (SO$_2$) feature at 4.0 $\mu$m is the best indicator of an exoVenus, but it is detectable in atmospheres with reduced CO$_2$ abundance.

\end{abstract}



\section{Introduction}
\label{intro}

Despite its currently uninhabitable surface conditions, interest in Venus has been growing significantly over the past decade. This is due in part to realizations regarding the importance of Venus to planetary habitability \citep{Kane2019}, whether it maintained a temperate climate up to $\sim$0.7~Gya \citep{Way2016,way2020}, or was unable to condense surface liquid water after its magma-ocean phase \citep{hamano2013,turbet2021}. The ambiguity regarding the evolution of Venus through time makes its current environment even more intriguing and further illustrates the need to understand why the climate of Venus may have diverged so drastically from Earth. The next generation of Venus missions such as the Deep Atmosphere Venus Investigation of Noble gases, Chemistry, and Imaging \citep[DAVINCI;][]{garvin2022revealing}, Venus Emissivity, Radio Science, InSAR, Topography, and Spectroscopy \citep[VERITAS;][]{cascioli2021determination}, and EnVision \citep{ghail2017envision} missions will be crucial for understanding Venus' history by constraining its water-loss history, determining the extent of geological activity, and providing insight into its interior structure.

A complimentary approach to studying possible evolutionary pathways of Venus is through the study of Venus-like exoplanets. Due to the intrinsic bias of the transit method toward smaller star--planet separations \citep{kane2008b}, the Kepler and Transiting Exoplanet Survey Satellite \citep[TESS;][]{Ricker2015} missions have discovered a plethora of terrestrial exoplanets with orbits in the Venus Zone \citep[VZ;][]{kane2014e,ostberg2019,ostberg2023demographics}. The TESS VZ planets are of particular interest as their host stars are much brighter than those of Kepler planets, making TESS planets more amenable to atmospheric spectroscopy with the James Webb Space Telescope (JWST) or other future facilities \citep{stassun2019revised,Louie2018}. Various studies have modelled the transmission spectra of potential exoVenuses and predicted the efficiency at which JWST could observe them  \citep[e.g.][]{lustig2019detectability,lustig2019mirage,lincowski2018evolved,ehrenreich2012transmission,way2023synergies}. Obtaining information from an exoVenus atmosphere will be a challenging endeavor however, as Venus-like clouds and hazes may prevent the detection of molecular species, or an atmosphere at all \citep[e.g. ][]{komacek2020clouds, fauchez2019impact, ehrenreich2006transmission, barstow2020unveiling}. In addition, it has been demonstrated that retrieval algorithms have difficulty discerning an Earth-like transmission spectrum from a Venus-like spectrum \citep{Barstow2016venus}. The inability to confidently identify an exoVenus or exoEarth will hinder efforts to learn the primary factors for Venus developing uninhabitable conditions, which is essential for understanding the circumstances which lead to the development of habitable worlds.

In this work we present an analysis of both hypothetical exoEarth and exoVenus transmission spectra with varying atmospheric CO$_2$ abundance to investigate potential pathways to differentiating the two planets. This entails determining the key spectral differences between an exoEarth and exoVenus and quantifying the detectability of the unique features in each planets' spectra. In Section~\ref{methods} we describe the climate model used to produce the exoEarth and exoVenus atmospheres, the process of simulating transmission spectra from the model output, and how we quantified the detectability of absorption features. Section~\ref{Results} goes into detail about the detectability of all major absorption features in the transmission spectra of both planet types, and how their detectability changes as a function of atmospheric CO$_2$ abundance. In Section~\ref{discussion} we discuss the features which are unique to each planets' spectra, the type of observing proposal that would be required to detect them, and the caveats of our analysis. Lastly, Section~\ref{conclusions} includes a summary of the main results and concluding remarks.


\section{Methods}
\label{methods}


\subsection{Modeling ExoEarth and ExoVenus Atmospheres} 

To generate model atmospheres for Venus-like and Earth-like conditions, we followed the methods and models of \citet{lincowski2018evolved} and Meadows et al. (2023, in prep.), respectively. As in those works, we used VPL Climate, a layer-by-layer, spectrum-resolving 1D climate model, which uses mixing length theory and latent heat exchange where appropriate to time-step a temperature profile to radiative-convective equilibrium \citep{robinson2018linearized,lincowski2018evolved}. The inputs required for VPL Climate are described in the respective model papers, and summarized below.

For radiative transfer, VPL Climate uses the Spectral Mapping Atmospheric Radiative Transfer (SMART) model, which is an accurate, spectrum-resolving radiative transfer model \citep{meadows1996ground,crisp1997thermal}, and has been successfully applied to both Earth \citep{crisp1997thermal,robinson2011earth} and Venus \citep{meadows1996ground,arney2014spatially,lincowski2021claimed}. SMART uses the Discrete Ordinate Radiative Transfer \citep[DISORT,][]{Stamnes1988,Stamnes2000} solver to calculate the radiation field. SMART uses LBLABC \citep{crisp1997absorption} to compute line-by-line absorption coefficients, and resolves both water (H$_2$O) and CO$_2$ line wings to 1000 cm$^{-1}$. The extensive Ames line list \citep{Huang2017} is used for CO$_2$ for the exoVenuses, HITEMP2010  \citep{HITEMP2010} is used for H$_2$O and CO for the exoVenuses, and HITRAN2016 \citep{HITRAN2016} is used for the remaining gases and for the exoEarths. LBLABC includes self and foreign line broadening. SMART uses laboratory-measured collision-induced absorption (CIA) for CO$_2$-CO$_2$, O$_2$-O$_2$, and N$_2$-N$_2$, and ultraviolet (UV) cross-sections as necessary (see \citealt{lincowski2018evolved} for a description of those data sources). For the exoEarths, moist convection is obtained through mixing length theory and latent heat release. Meadows et al. (2023, in prep.) showed that this, coupled with the photochemistry model, faithfully reproduces the global conditions of Earth’s atmosphere. In both cases, VPL Climate was coupled to the photochemical component of Atmos, as described in \citet{lincowski2018evolved} and Meadows et al. (2023, in press.).

Aerosols and clouds are included in the VPL Climate simulations. For the exoEarths, we used standard Earth stratocumulus (water) and cirrus (water-ice) clouds to obtain the global surface temperature. The stratocumulus clouds were defined to be between 0.827--0.900 bars, and the cirrus clouds are between 0.257--0.331 bars. The clouds were described in Meadows et al. (2023, in press.) and derive from Earth atmospheric observations \citep{robinson2011earth}. For exoVenuses, we generate optical depths and aerosol properties converted from monodisperse, layer-by-layer sulfuric acid (H$_2$SO$_4$) aerosols generated in the photochemical code. These aerosols vary in particle size and H$_2$SO$_4$ fraction. A more detailed description of the H$_2$SO$_4$ aerosols can be found in \citet{lincowski2018evolved}.

\begin{figure*}
\includegraphics[width = \textwidth]{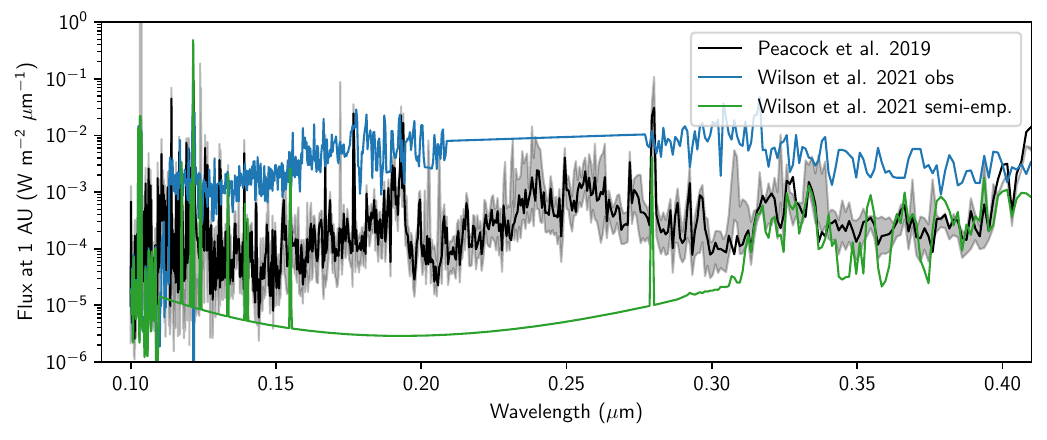}
\caption{A comparison of three TRAPPIST-1 spectra in the UV. The black line is the average of three spectra simulated by \citet{Peacock2019}, and is the spectrum used in the VPL Climate simulations. Both the blue and green lines are spectra from \citet{wilson2021mega}, and were produced using observational data and a semi-empirical model, respectively.}
\label{fig:T1SED}
\end{figure*}

The exoEarths and exoVenuses were both modeled as 1~M$_\oplus$, 1~R$_\oplus$ planets, near the runaway greenhouse limit for TRAPPIST-1. We chose TRAPPIST-1 as the host star since the majority of exoplanets are being discovered around cooler M-type stars. For both photochemical and climate modeling, we used a version of the \citet{Peacock2019} synthetic spectrum calibrated to GALEX near--ultraviolet (NUV) and far--ultraviolet (FUV) fluxes. Figure~\ref{fig:T1SED} shows the \citep{Peacock2019} spectrum in comparison to two spectra from \citet{wilson2021mega}, in units of specific flux converted to SI units at 1 AU. The \citet{Peacock2019} spectrum, which had spurious lines removed, is an average of the three spectra simulated by \citet{Peacock2019}, the range of which are shaded in grey. We also plot the Hubble Space Telescope (HST) Cosmic Origins Spectrograph (COS) observations published by \citet{wilson2021mega} in blue, which is missing a segment of data between approximately 0.2--0.3 $\mu$m. For planetary modeling, as we conduct here, \citet{wilson2021mega} recommended using their semi-empirical model, which is plotted in green. The semi-empirical model replaces essentially the entire critical NUV and FUV flux range with a polynomial fit. In our opinion, this is insufficient for planetary modeling, given that \citet{Peacock2019} conducted a careful UV spectral model, which covers this range. As can be seen in Figure~\ref{fig:T1SED}, the \citet{Peacock2019} model lies largely in between the low flux of the semi-empirical model and the flux observed by HST COS. \citet{Peacock2019} noted their models were representative of upper limits as observed by GALEX in the NUV and FUV bands, the spectrum shown is still lower than the HST COS observations. \citet{wilson2021mega} had recommended to not use the observed spectrum, due to low S/N of M dwarf UV observations, particularly for TRAPPIST-1. Although individual strong line fits may be poor according to \citet{wilson2021mega}, the integrated flux throughout the FUV and NUV is important for planetary photochemical modeling. Therefore, we use the averaged \citet{Peacock2019} spectrum in our climate models.

\begin{figure*}
\includegraphics[width = \textwidth]{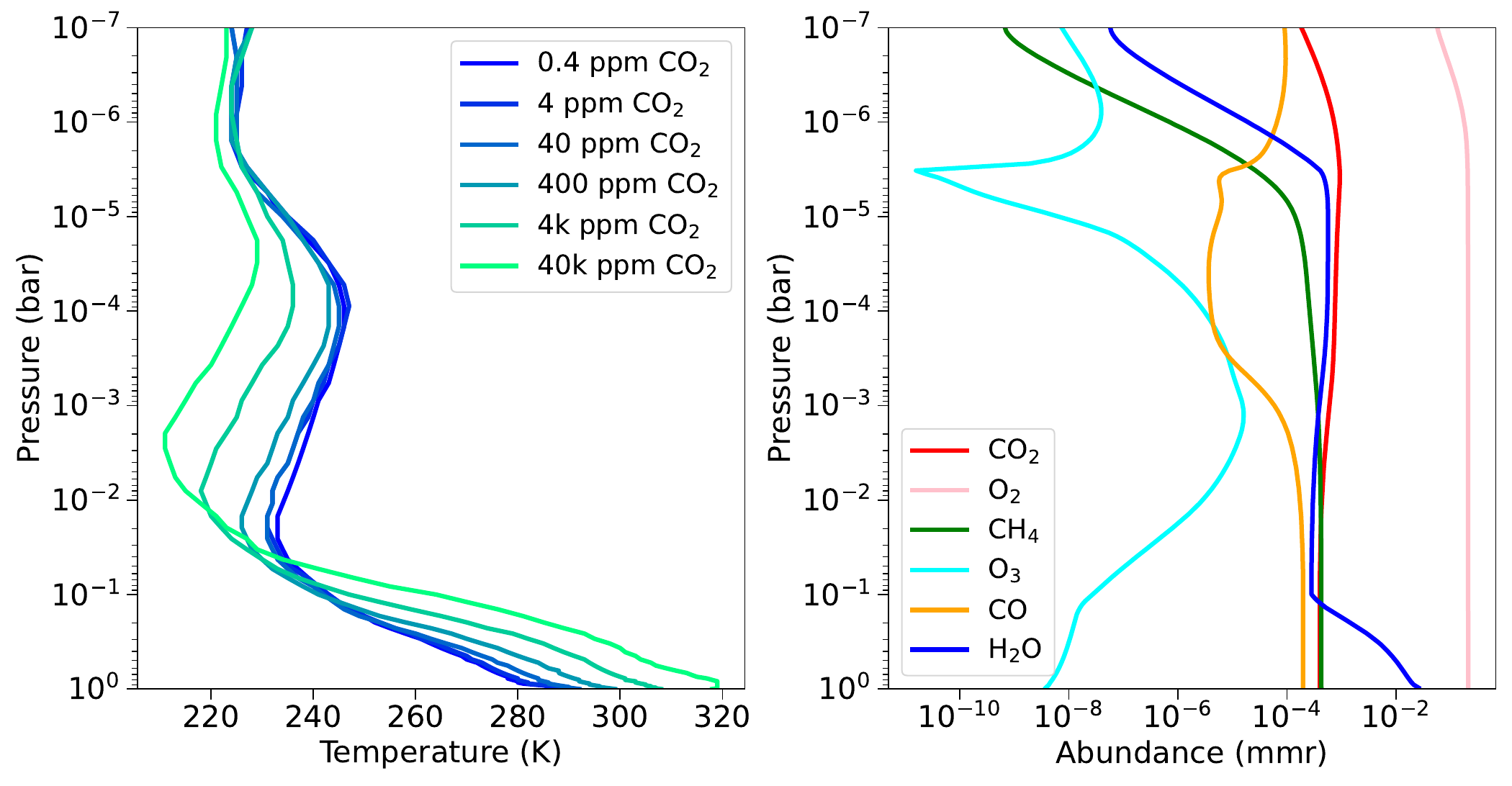}
\caption{The TP profiles for the 6 exoEarth atmospheres (left panel) and the chemical abundance profiles for the 400 ppm exoEarth (right panel).}
\label{EarthTP}
\end{figure*}

\begin{figure*}
\includegraphics[width = \textwidth]{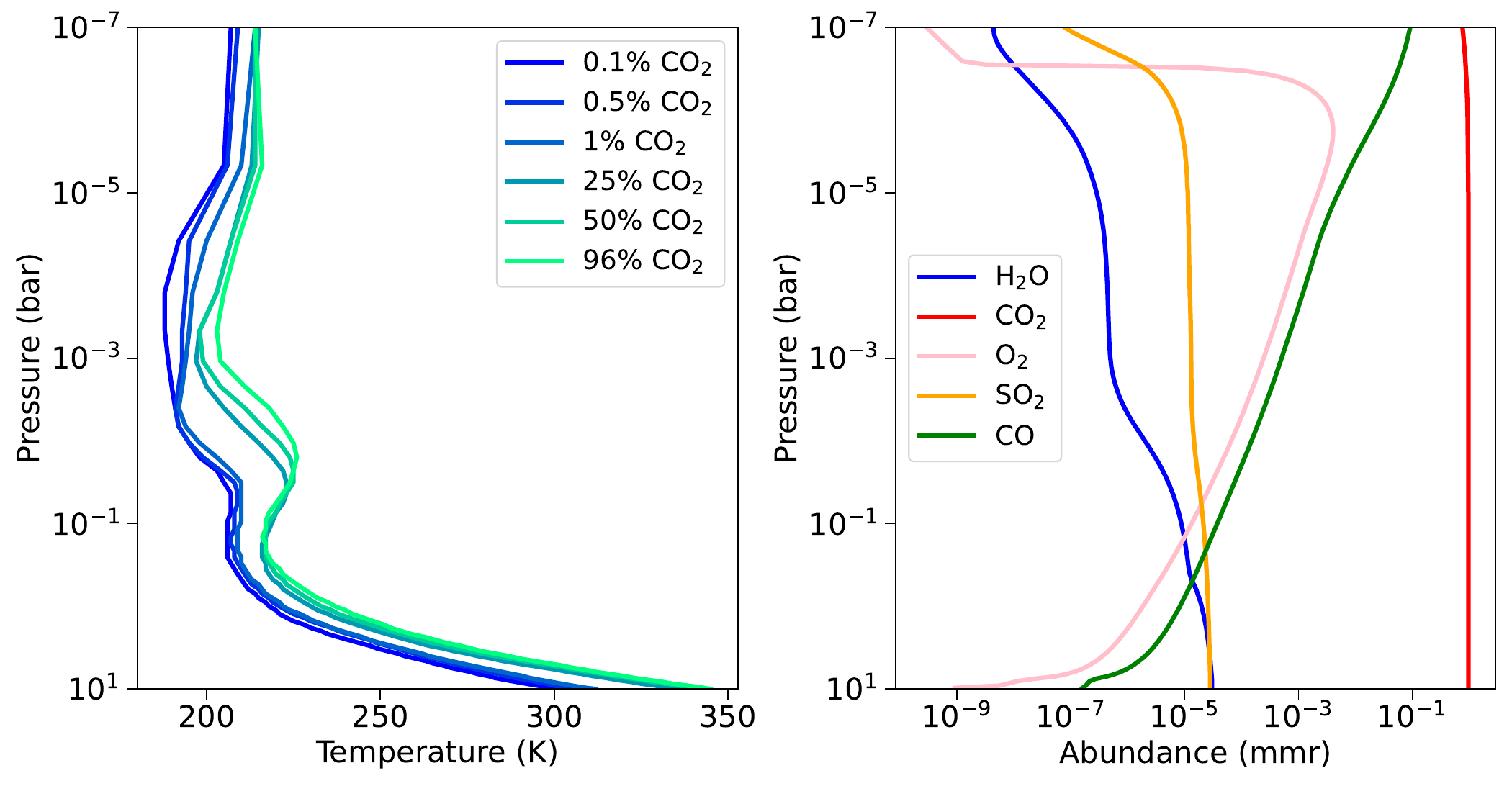}
\caption{The TP profiles for the 6 exoVenus atmospheres (left panel) and the chemical abundance profiles for the 96\% CO$_2$ exoVenus (right panel).}
\label{VenusTP}
\end{figure*}

To accommodate varying CO$_2$ levels, N$_2$ was exchanged for CO$_2$ to keep the surface pressure as 1~bar for the exoEarths, and 10~bar for the exoVenuses. The exoEarth atmospheres are composed of H$_2$O, CO$_2$, ozone (O$_3$), nitrous oxide (N$_2$O), carbon monoxide (CO), methane (CH$_4$), oxygen (O$_2$), sulfur dioxide (SO$_2$), carbonyl sulfide (OCS), ethane (C$_2$H$_6$), dimethyl sulfide (C$_2$H$_6$S), hydrogen chloride (HCl), and chloromethane (CH$_3$Cl). The exoVenuses have atmospheres made of H$_2$O, CO$_2$, CO, SO$_2$, OCS, O$_2$, HCl, and H$_2$SO$_4$. The exoEarths used the average surface albedo for Earth \citep{meadows2018habitability} and the exoVenuses used a basalt surface \citep{lincowski2018evolved}. The resulting temperature-pressure (TP) profiles for all exoEarth and exoVenus cases are shown in Figures \ref{EarthTP} and \ref{VenusTP}.

\subsection{PSG and PandExo}

The transmission spectra for the exoEarths and exoVenuses were modeled using the Planetary Spectrum Generator \citep[PSG; ][]{villanueva2018planetary}. PSG is a publicly available online interface that couples radiative transfer models with planetary and spectral databases. PSG requires input which describes the host star, planet, orbital configuration, and the observing instrument. The host star used was TRAPPIST-1, and all exoEarth and exoVenus were defined to have 1 M$_\oplus$, 1 R$_\oplus$, and a semi-major axis on the runaway greenhouse boundary \citep[0.02393 AU;][]{Kopparapu2013,kopparapu2014habitable}. The observing instrument used in the PSG simulations was JWST NIRSpec PRISM, which has a bandpass of 0.6--5.3 $\mu$m. The atmospheres for the exoEarths and exoVenuses were defined using the TP and molecular abundance profiles from the Atmos simulations. Both cloudy and clear-sky transmission spectra were produced for each planet. We did not include emission spectra since secondary eclipse observations are typically used for detecting the presence of atmospheres and not specific molecules, as was done with TRAPPIST-1 b \citep{greene2023thermal} and TRAPPIST-1 c \citep{zieba2023no}. 

\begin{figure*}
\includegraphics[width=\textwidth]{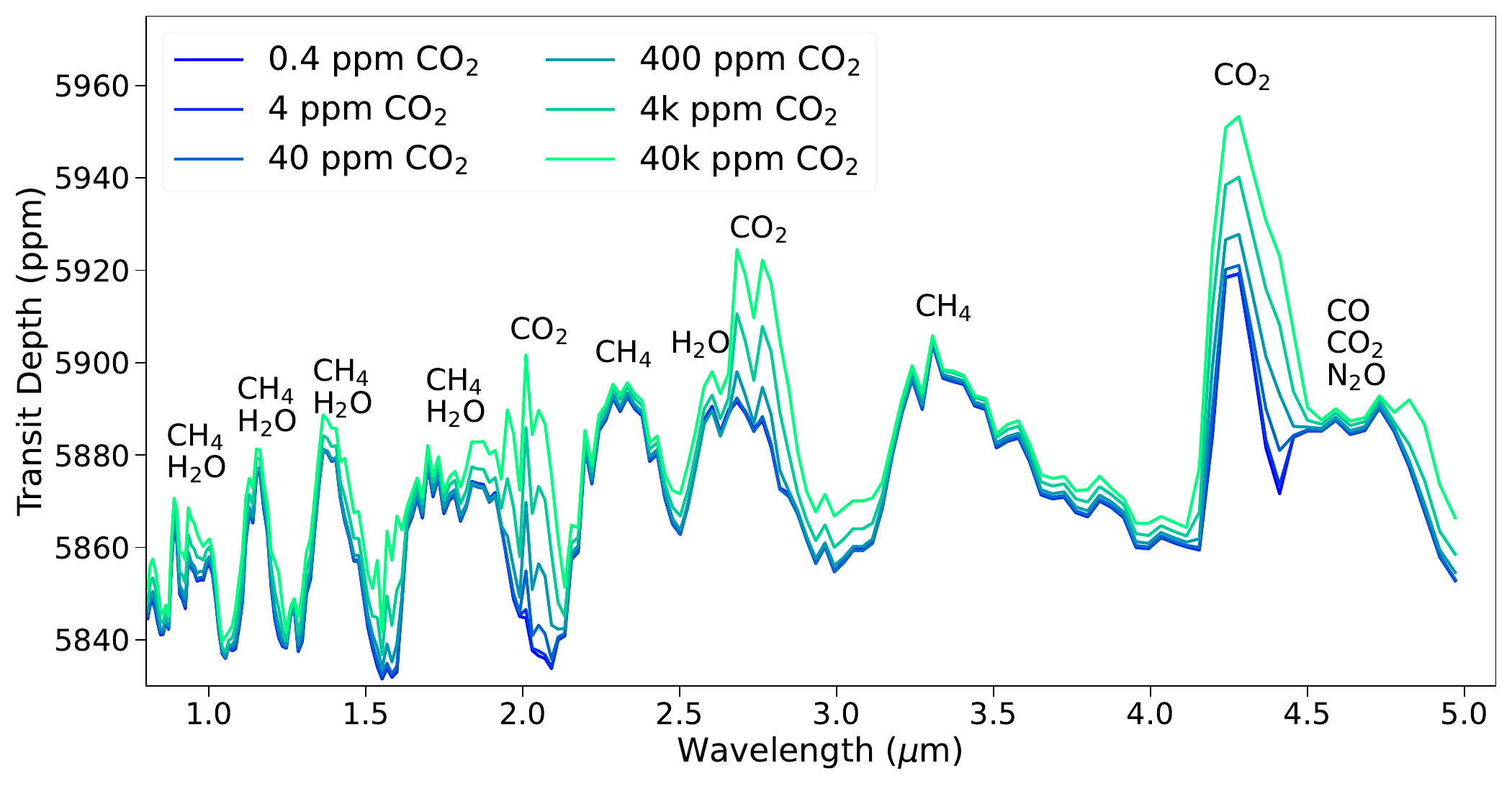}
\includegraphics[width=\textwidth]{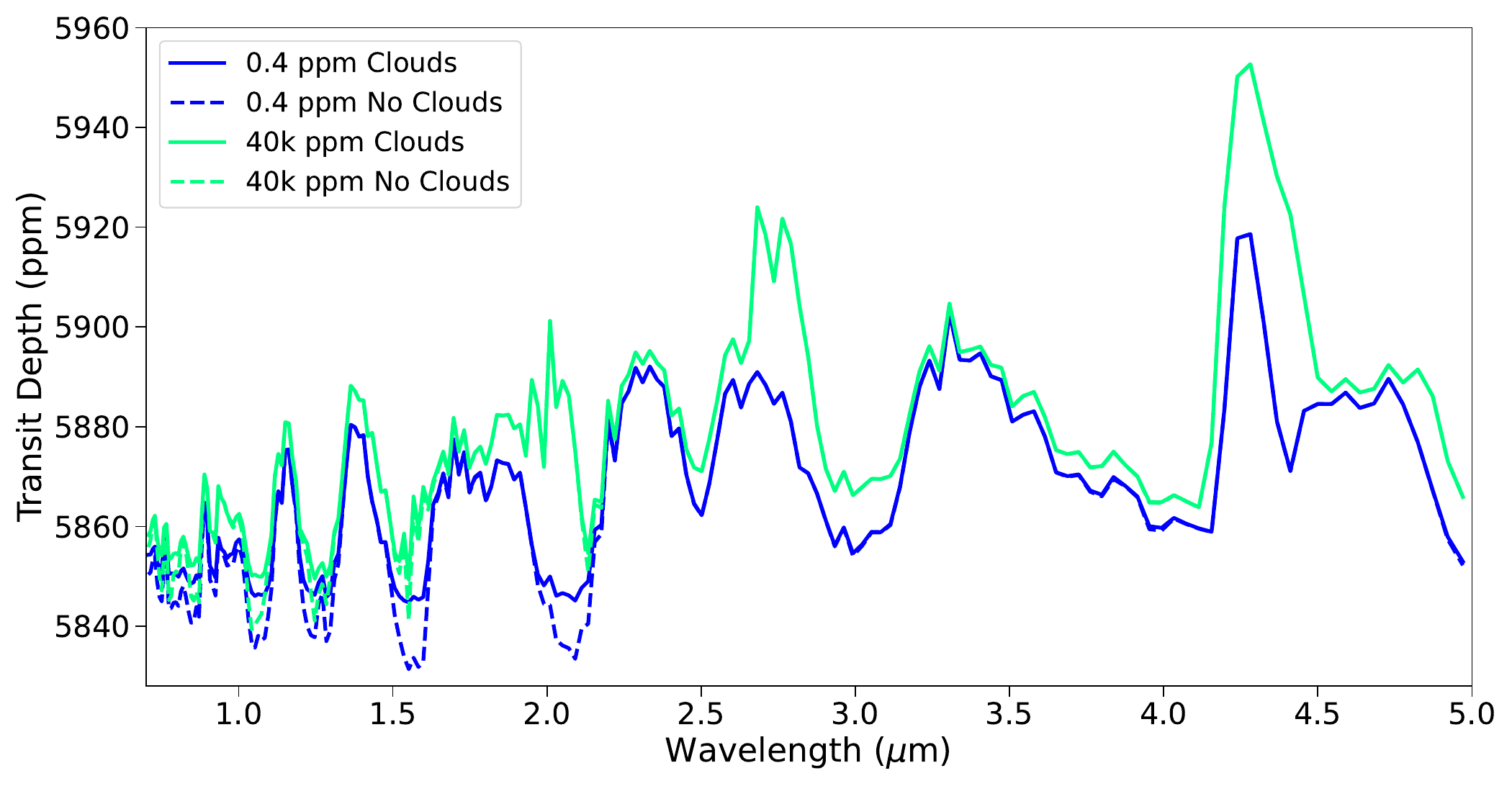}
\caption{The transmission spectra of the 6 clear-sky TRAPPIST-1 exoEarths from the VPL Climate simulations (upper panel), and transmission spectra with and without clouds for the 0.4 and 40k ppm CO$_2$ exoEarths. Absorption features are labelled with the molecules which contribute the most absorption. Features labelled with multiple molecules consist of more than one primary absorber.}
\label{fig:T1E_AllSpec}
\end{figure*}

\begin{figure*}
\includegraphics[width=\textwidth]{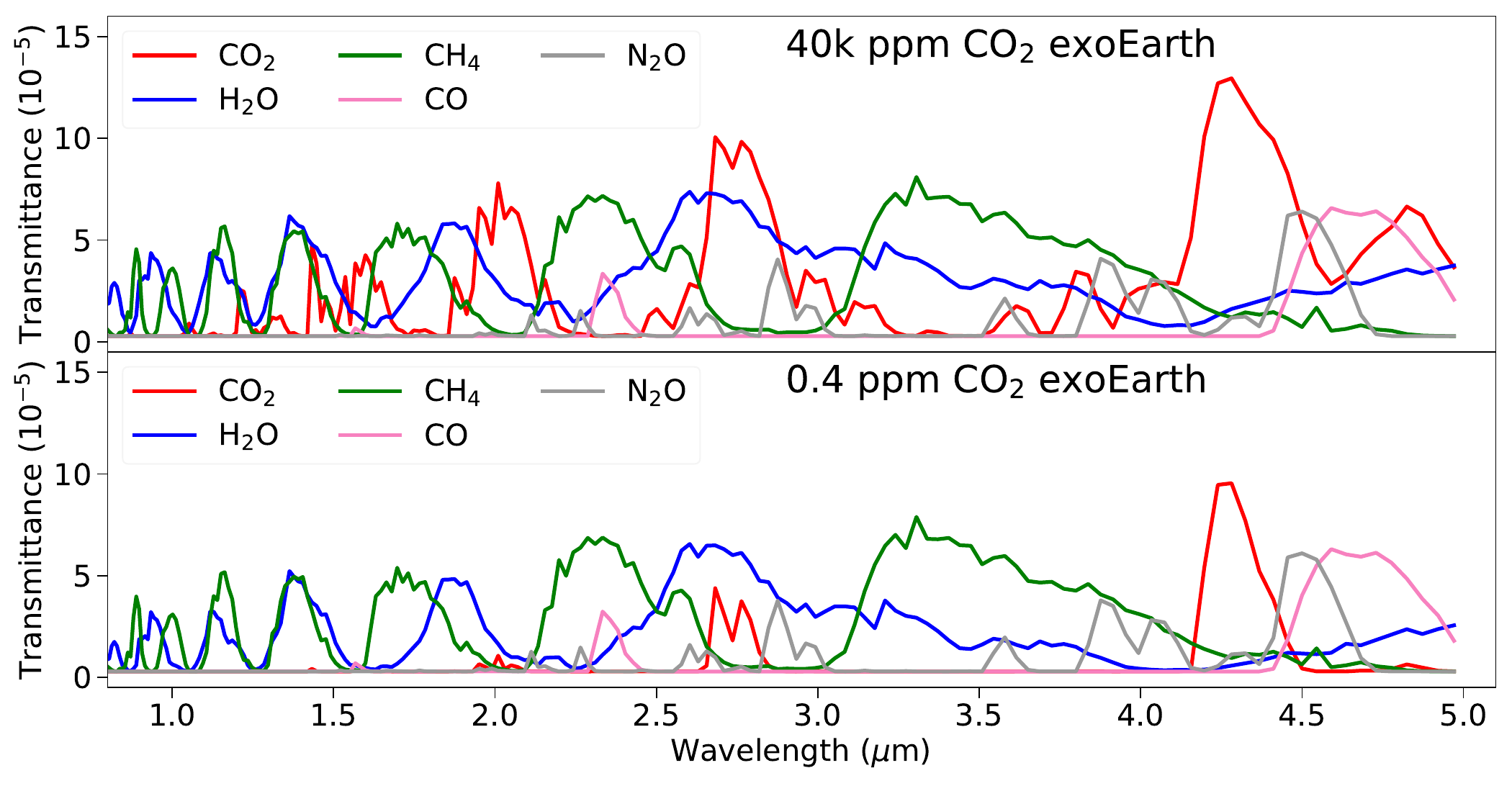}
\caption{The transmittance of each molecular species in the 40k ppm (upper panel) and 0.4 ppm (lower panel) CO$_2$ exoEarth atmospheres. Absorption by CO$_2$ is much stronger in the 40k ppm exoEarth, but all other molecular absorption remains constant in both cases.}
\label{fig:T1E_Transmittance}
\end{figure*}

The upper panel of Figure~\ref{fig:T1E_AllSpec} shows the PSG transmission spectra for all 6 exoEarths assuming clear-skies. Increasing CO$_2$ abundance has little effect on the exoEarth spectra until the abundance reaches 4k ppm CO$_2$. In the 4k and 40k ppm CO$_2$ spectra, the 2.0 and 2.7 $\mu$m CO$_2$ features become visible. The 4.3$\mu$m CO$_2$ feature is visible in every CO$_2$ case, but only increases in size in the 4k and 40k ppm cases. All other features are primarily composed of either H$_2$O or CH$_4$, and their size remains constant in all cases. The modelled exoEarth atmospheres have greater CH$_4$ absorption than Earth because the spectral energy distribution of TRAPPIST-1 enhances the production of CH$_4$ \citep{meadows2018habitability}. Other molecules with smaller contributions to the exoEarth absorption features include CO and N$_2$O (Figure~\ref{fig:T1E_Transmittance}). The lower panel of Figure~\ref{fig:T1E_AllSpec} displays the effect of water and water-ice clouds on the 0.4 and 40k ppm exoEarth transmission spectra. Since the clouds are located at low altitudes they have little effect on the spectra, and only slightly raise the continuum at wavelengths less than 2.3 $\mu$m.

\begin{figure*}
\includegraphics[width=\textwidth]{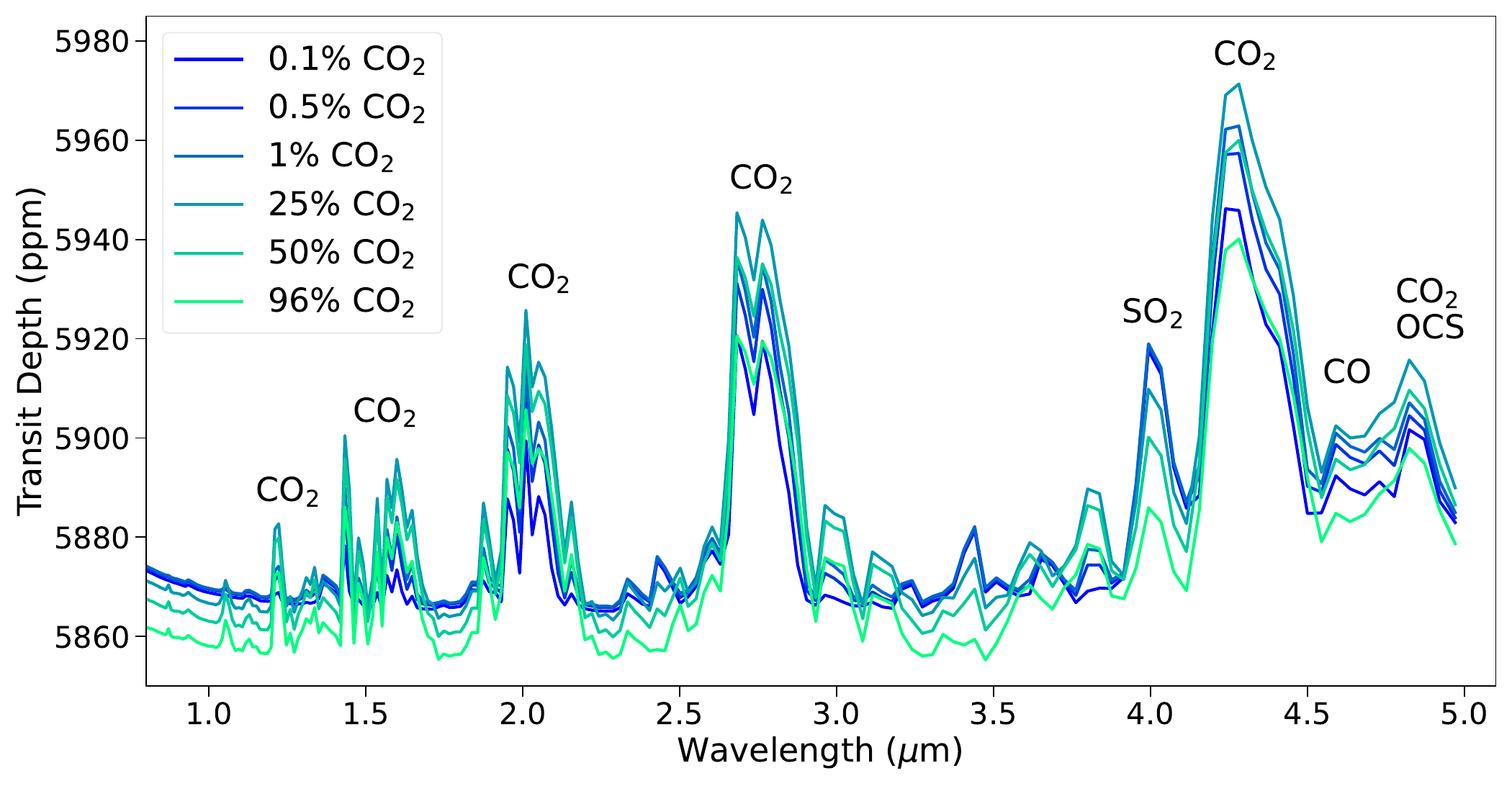}
\includegraphics[width=\textwidth]{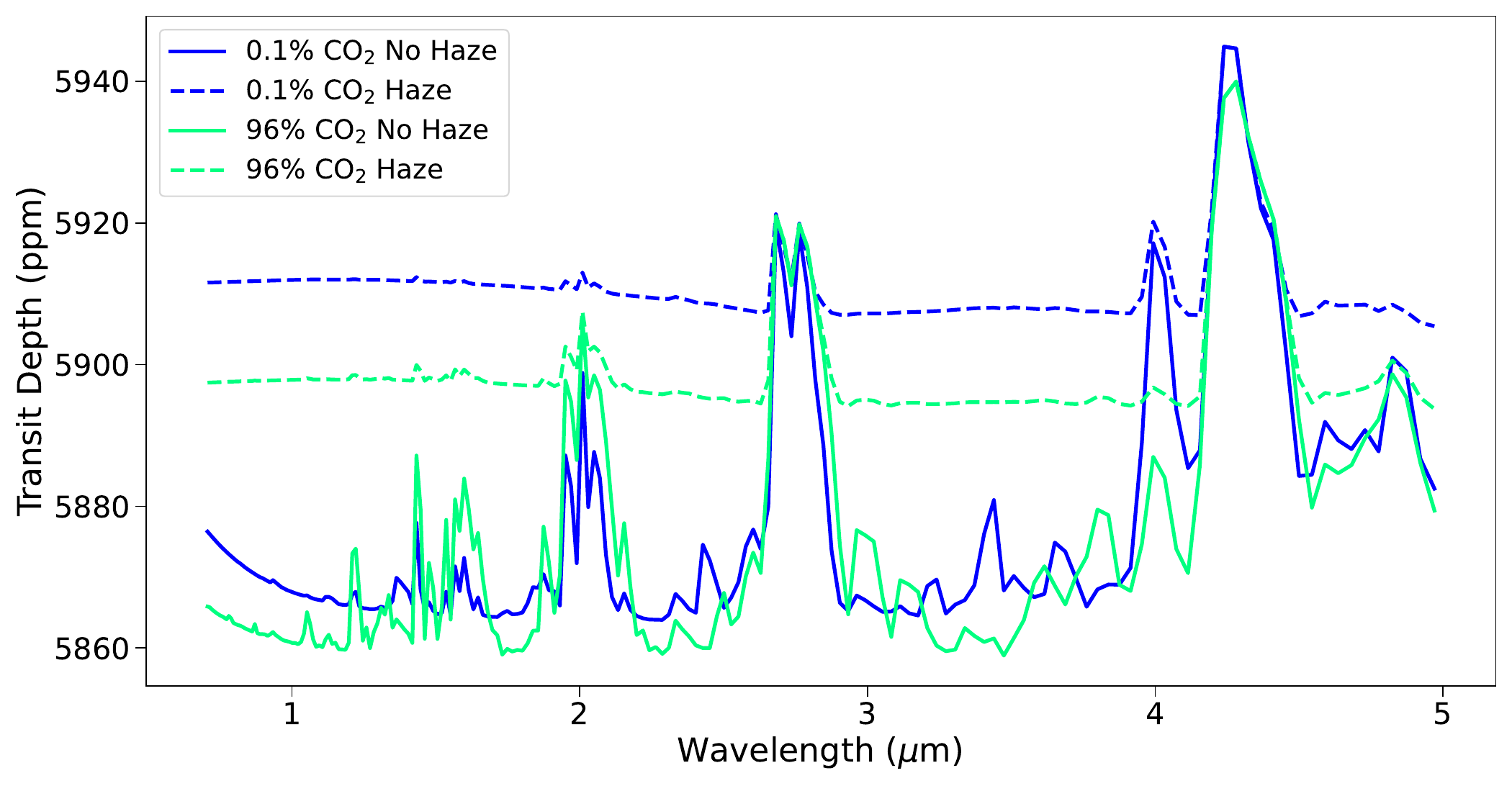}
\caption{The transmission spectra of the 6 clear-sky TRAPPIST-1 exoVenuses from the VPL Climate simulations (upper panel), and transmission spectra with and without haze for the  and 0.1\% and 96\% CO$_2$ exoVenuses (lower panel). Absorption features are labelled with the molecules which contribute the most absorption. Features labelled with multiple molecules consist of more than one primary absorber.}
\label{fig:T1V_AllSpec}
\end{figure*}

\begin{figure*}
\includegraphics[width=\textwidth]{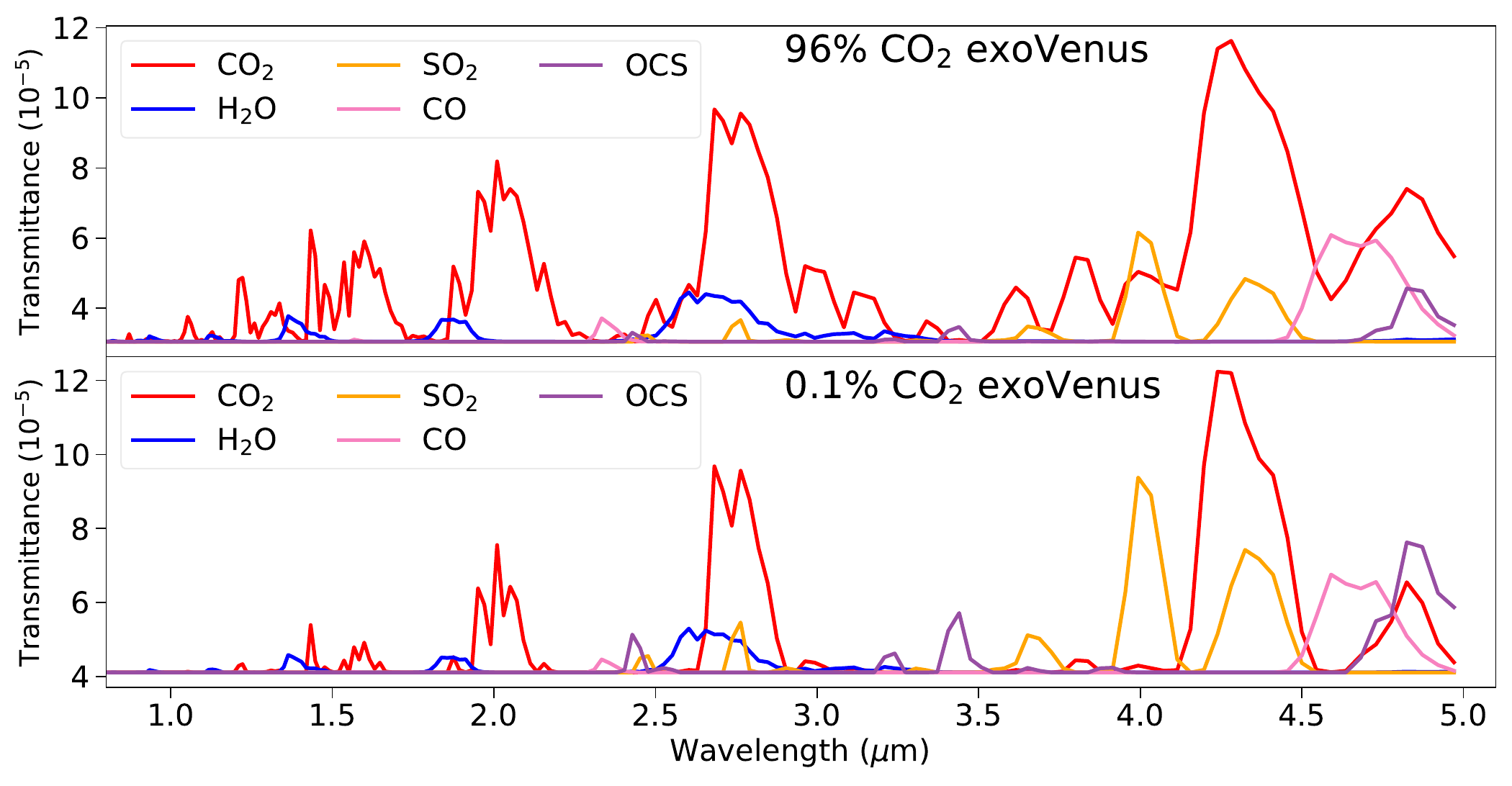}
\caption{The transmittance of each molecular species in the 96\% (upper panel) and 0.1\% (lower panel) CO$_2$ exoVenus atmospheres. The decreased CO$_2$ absorption in the 0.1\% CO$_2$ case allows for more absorption by the SO$_2$ feature at 4.0 $\mu$m.}
\label{fig:T1V_Transmittance}
\end{figure*}

The exoVenus spectra have far more variation with changing CO$_2$ abundance since majority of its features are composed of only CO$_2$ absorption (Figure~\ref{fig:T1V_AllSpec}). The only other prominent molecular absorption is caused by sulfur dioxide (SO$_2$) at 4.0 $\mu$m and CO at 4.6 $\mu$m. In the clear-sky exoVenus spectra, the SO$_2$ feature is only visible in lower CO$_2$ cases. In the higher CO$_2$ atmospheres the SO$_2$ feature is concealed because of increased atmospheric scale height and increased CO$_2$ absorption (Figure~\ref{fig:T1V_Transmittance}). Larger CO$_2$ abundance does not equate to larger CO$_2$ features in the 50\% and 96\% CO$_2$ cases due to an increase in scale height which compresses the atmosphere and reduces the size of absorption features. As a result, the 50\% and 96\% CO$_2$ exoVenus spectra have features similar in size to that of the 0.1\% CO$_2$ spectrum. The scale height and CO$_2$ absorption in the 25\% CO$_2$ spectrum are optimal, allowing it to have the largest features of the 6 exoVenuses. 

Unlike the clouds in the exoEarth spectra, the H$_2$SO$_4$ haze in the exoVenus spectra significantly impacts the size of absorption features (Figure~\ref{fig:T1V_AllSpec}). The haze causes the CO$_2$ feature at 1.7 $\mu$m to be entirely muted, and the CO$_2$ feature at 2.0 $\mu$m is only slightly visible. In the 0.1\% CO$_2$ cloudy spectrum, both the 2.7 $\mu$m CO$_2$ feature and 4.0 $\mu$m SO$_2$ feature are reduced to a height of about 10 ppm. The 4.3 $\mu$m CO$_2$ is more resilient to the haze, and maintains a height of about 20 and 40 ppm in the 0.1\% and 96\% CO$_2$ spectra, respectively.

The transmission spectra in Figure~\ref{fig:T1V_AllSpec} and Figure~\ref{fig:T1V_AllSpec} were used as inputs for PandExo \citep{batalha2017pandexo} to simulate observations with JWST NIRSpec PRISM. PandExo utilizes the Space Telescope Science Institute's (STSI) Exposure Time Calculator (ETC), and Pandeia \citep{pontoppidan2016pandeia}, to model instrumental and background noise sources. The physical and orbital parameters for the planet and host star were defined in PandExo to be the same values used for PSG. The atmosphere of the host star is generated by PandExo using the Phoenix Stellar Atlas \citep{husser2013new}, and each planets' atmosphere was defined by their PSG transmission spectrum. 

\begin{figure*}
\includegraphics[width=\textwidth]{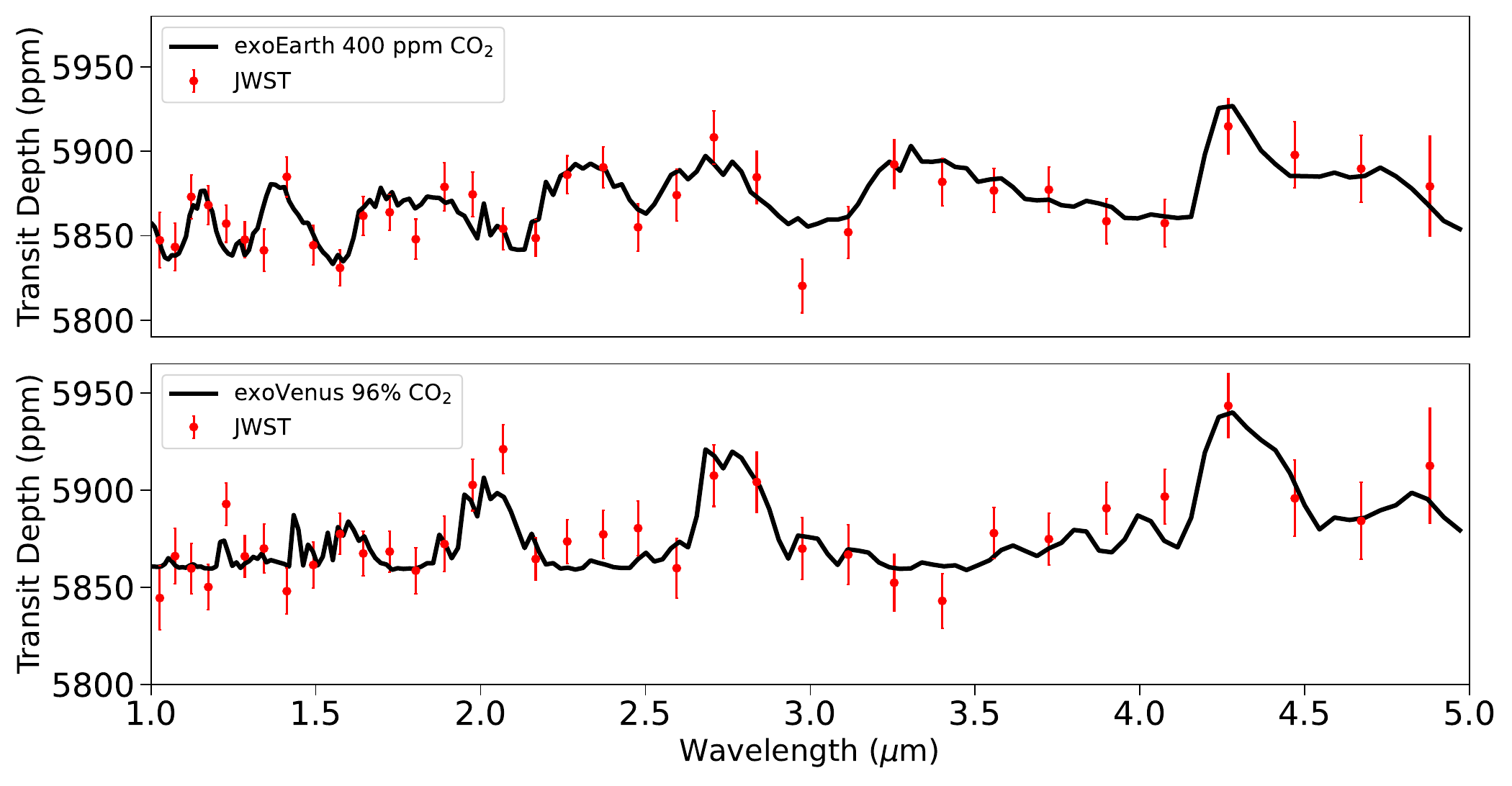}
\caption{Simulated JWST NIRSpec PRISM spectroscopy data from 15 transit observations of the 400 ppm CO$_2$ exoEarth (top) and 96\% CO$_2$ exoVenus (bottom), both without clouds or hazes. The JWST data was binned down to a resolving power of $R=12$ for easier visibility.}
\label{fig:PandexoExample}
\end{figure*}

NIRSpec PRISM was defined to have a saturation level of 80\% a full well \citep{Greene2016}, and the noise floor was set to 5 ppm to reflect the noise seen in recent NIRSpec observations \citep{lustig2023jwst}. The ratio of in-transit to out of transit observing time was set to 1, making a single transit observation equivalent to 1.75 hours of observation, assuming no overhead time. We use the SUB512 subarray with 6 groups per integration and the native binning of NIRSpec PRISM ($R$ = 100). An example of PandExo simulated JWST data assuming 15 transit observations of both the 400 ppm CO$_2$ exoEarth and 96\% CO$_2$ exoVenus are shown in Figure~\ref{fig:PandexoExample}. The simulated data has worse uncertainty and spectral resolution at longer wavelengths because of the sensitivity of NIRSpec PRISM. This causes the JWST data to resolve less of the 4.3 $\mu$m feature than the features at shorter wavelengths, despite the 4.3 $\mu$m feature being larger.

\subsection{Determining the S/N of Absorption Features}
\label{sec:SNcalcs}

We calculated the S/N of prominent absorption features in the exoEarth and exoVenus transmission spectra in order to quantify their detectability. The $\chi^2$ approach used by \citet{lustig2019detectability} was adopted for our S/N calculations:

\begin{equation}
    S/N = \sqrt{ \sum_{i}^{N_{\lambda_i}} \left ( \frac{y_i - y_{cont}}{\sigma_i} \right )^2 }
\label{SNEquation}
\end{equation}

In Equation \ref{SNEquation}, $y_i$ is the i'th y value of the model spectrum within the wavelength range of a feature, $\sigma_i$ is the corresponding uncertainty of the simulated JWST data from PandExo, and $y_{cont}$ is the continuum which we defined as the minimum value of the entire spectrum. 

The S/N of a given feature is first calculated for a single transit observation, and a scaling relationship is used to interpolate the S/N to up to 100 transits. We then determined the number of transits required for a feature to reach a S/N $\geq$ 5, which is the detectability threshold used in previous studies \citep[e.g.]{lustig2019detectability,pidhorodetska2020detectability,felton2022role}. If the threshold is not reached within 100 transit observations then the feature is determined to be undetectable. Since more complex retrieval models will be required to confirm the presence of molecular species from actual JWST data, we acknowledge that the values for S/N and number of transits that we report are to be considered upper and lower limits, respectively.

We chose to focus on 10 features in the exoEarth spectra and 7 in the exoVenus spectra, which are labeled in Figure~\ref{fig:T1E_AllSpec} and Figure~\ref{fig:T1V_AllSpec}. For reference, the defined wavelength ranges and molecular absorbers for every feature in the exoEarth and exoVenus spectra are listed in Table~\ref{tab:T1E} and Table~\ref{tab:T1V}, respectively. The number of transits required to detect every feature in the exoEarth and exoVenus spectra were calculated using Equation~\ref{SNEquation}, and the results are discussed in Section \ref{Results}


\section{Results}
\label{Results}

Using Equation~\ref{SNEquation} we determined the number of transit observations required to detect the 10 absorption features in the exoEarth transmission spectra and 7 features in the exoVenus spectra. Shown in Table~\ref{tab:T1E} are the number of transits required for the features in the 6 clear-sky exoEarth cases to reach the S/N threshold. The CH$_4$ feature at 0.9 $\mu$m is the smallest of the features and therefore requires the most time to detect. The size of the CH$_4$ and H$_2$O features did not change with increasing CO$_2$ (Figure~\ref{fig:T1E_AllSpec}), causing the number of transits required to detect to remain relatively constant across all CO$_2$ cases. The CO$_2$ features at 2.0, 2.7, and 4.3 $\mu$ only grew in size when CO$_2$ was increased to 4k and 40k ppm. This caused the number of transits required for detection of these features to be relatively unchanged in all but the 4k and 40k ppm CO$_2$ exoEarth cases. The CO$_2$ feature at 2.0 $\mu$m is undetectable when CO$_2$ abundance is less than 4k ppm. The CH$_4$ features at 1.7, 2.3, and 3.5 $\mu$m are are the most consistently detectable features in the exoEarth spectra as they can be detected in at least 9, 7, and 6 transits, respectively. Table~\ref{tab:T1E_Cloudy} lists the number of transits required to detect features in the cloudy exoEarth transmission spectra. Since clouds have a negligible effect on the exoEarth spectra (Figure~\ref{fig:T1E_AllSpec}), the required transits are the same for both the cloudy and clear-sky spectra.

\begin{deluxetable*}{ c c c c c c c c }
\tabletypesize{\scriptsize}
\tablecaption{Detecting Clear-Sky ExoEarth Absorption Features}
\tablehead{\colhead{Feature} & \colhead{Wavelengths} & \multicolumn{6}{c}{Transits Required for Detection} \\
\cline{3-8}
\colhead{} & \colhead{$\mu$m} & \colhead{0.4 ppm} & \colhead{4 ppm} & \colhead{40 ppm} & \colhead{400 ppm} & \colhead{4k ppm} & \colhead{40k ppm} }
\startdata
CH$_4$ + H$_2$O & 0.85--1.06 & 81 & 81 & 82 & 84 & 86 & 84\\
CH$_4$ + H$_2$O & 1.08--1.24 & 25 & 25 & 25 & 25 & 26 & 25\\
CH$_4$ + H$_2$O & 1.28--1.58 & 17 & 17 & 17 & 18 & 17 & 15\\
CH$_4$ + H$_2$O & 1.59--2.0 & 9 & 9 & 9 & 9 & 9 & 8 \\
CO$_2$ & 2.0--2.13 & - & - & - & - & 49 & 26 \\
CH$_4$ & 2.15--2.48 & 7 & 7 & 7 & 7 & 8 & 8 \\
H$_2$O + CO$_2$ & 2.5--3.0 & 11 & 11 & 12 & 11 & 10 & 8 \\
CH$_4$ & 3.14--4.03 & 6 & 6 & 6 & 6 & 7 & 7 \\
CO$_2$ & 4.11--4.45 & 18 & 17 & 16 & 14 & 11 & 9 \\
CO + N$_2$O & 4.46--5.0 & 28 & 28 & 28 & 29 & 30 & 29 \\
\enddata
\tablecomments{The dominant molecules and wavelength ranges for the 10 absorption features in the exoEarth spectra without clouds. The last 6 columns of the table list the number of transits needed to detect a given feature in each of the 6 exoEarth CO$_2$ cases. Dashes indicate the feature was unable to be detected in less than 100 transit observations.}
\label{tab:T1E}
\end{deluxetable*}

\begin{deluxetable*}{ c c c c c c c c }
\tabletypesize{\scriptsize}
\tablecaption{Detecting Cloudy ExoEarth Absorption Features}
\tablehead{\colhead{Feature} & \colhead{Wavelengths} & \multicolumn{6}{c}{Transits Required for Detection} \\
\cline{3-8}
\colhead{} & \colhead{$\mu$m} & \colhead{0.4 ppm} & \colhead{4 ppm} & \colhead{40 ppm} & \colhead{400 ppm} & \colhead{4k ppm} & \colhead{40k ppm} }
\startdata
CH$_4$ + H$_2$O & 0.85--1.06 & 81 & 82 & 82 & 84 & 86 & 84\\
CH$_4$ + H$_2$O & 1.08--1.24 & 25 & 25 & 25 & 25 & 26 & 25\\
CH$_4$ + H$_2$O & 1.28--1.58 & 17 & 17 & 17 & 18 & 17 & 15\\
CH$_4$ + H$_2$O & 1.59--2.0 & 9 & 9 & 9 & 9 & 9 & 8 \\
CO$_2$ & 2.0--2.13 & - & - & - & - & 49 & 26 \\
CH$_4$ & 2.15--2.48 & 7 & 7 & 7 & 7 & 8 & 8 \\
H$_2$O + CO$_2$ & 2.5--3.0 & 11 & 11 & 12 & 11 & 10 & 8 \\
CH$_4$ & 3.14--4.03 & 6 & 6 & 6 & 6 & 7 & 7 \\
CO$_2$ & 4.11--4.45 & 18 & 17 & 16 & 14 & 11 & 9 \\
CO + N$_2$O & 4.46--5.0 & 28 & 28 & 28 & 29 & 30 & 29 \\
\enddata
\tablecomments{The dominant molecules and wavelength ranges for the 10 absorption features in the exoEarth spectra with clouds. The last 6 columns of the table list the number of transits needed to detect a given feature in each of the 6 exoEarth CO$_2$ cases. Dashes indicate the feature was unable to be detected in less than 100 transit observations.}
\label{tab:T1E_Cloudy}
\end{deluxetable*}

The number of transits required for the absorption features in the clear-sky exoVenus spectra to reach the S/N threshold are shown in Table \ref{tab:T1V}. The 25\% CO$_2$ case requires the least amount of transits to detect all CO$_2$ features, whereas the 0.5\% and 1\% CO$_2$ atmospheres are the best opportunity to detect SO$_2$. The large CO$_2$ feature at 4.3 $\mu$ requires a similar amount of transits to be detected in the 96\% and 0.1\% CO$_2$ cases because the increased scale height of the 96\% CO$_2$ atmosphere reduces the features size. The CO$_2$ features at 2.0 and 2.7 $\mu$m require far less transits to be detected in the 96\% CO$_2$ case than in the 0.1\% CO$_2$ case however, since these features are only prominent in atmospheres with high CO$_2$ abundance. The SO$_2$ feature at 4.0 $\mu$m is the only feature that is not also present in the exoEarth spectra, however it requires a minimum of 47 transits to be detected, and becomes undetectable in the 96\% CO$_2$ case. Alternatively, the CO$_2$ feature at 1.6 $\mu$m does not become detectable until there is at least 25\% atmospheric CO$_2$. The smallest CO$_2$ feature at 1.3 $\mu$m cannot be detected in less than 100 transit observations for any of the 6 exoVenuses. 

Table~\ref{tab:T1V_Cloudy} lists the amount of transits required to detect the features in the hazy exoVenus spectra. The haze prevents all but the 2.7 and 4.3 $\mu$ CO$_2$ features from being detected in any of the 6 exoVenus spectra. The 4.3 $\mu$ feature is detectable in as little as 33 transits in the 25\% CO$_2$ case, but can also become undetectable in the lowest CO$_2$ case. The 2.7 $\mu$m feature would require a minimum of 66 hours to detect. The haze concealing SO$_2$ absorption is particularly significant given that it is the only feature unique to the exoVenus spectra, and leads to ambiguity when using retrieval algorithms to derive surface conditions from the spectrum \citep{Barstow2016venus}.

\begin{deluxetable}{ c c c c c c c c }
\tabletypesize{\scriptsize}
\tablecaption{Detecting Clear-Sky ExoVenus Absorption Features}
\tablehead{\colhead{Feature} & \colhead{Wavelengths} &
\multicolumn{6}{c}{Transits Required for Detection} \\
\cline{3-8}
\colhead{} & \colhead{$\mu$m} & \colhead{0.1\%} & \colhead{0.5\%} & \colhead{1\%} & \colhead{25\%} & \colhead{50\%} & \colhead{96\%} }
\startdata
CO$_2$ & 1.16--1.4 & - & - & - & - & - & -   \\
CO$_2$ & 1.4--1.8 & - & - & - & 32 & 44 & 55   \\
CO$_2$ & 1.83--2.25 & 89 & 39 & 30 & 15 & 19 & 25 \\
CO$_2$ & 2.5--3.08 & 30 & 19 & 16 & 11 & 13 & 18 \\
SO$_2$ & 3.9--4.11 & 52 & 47 & 47 & 63 & - & - \\
CO$_2$ & 4.11--4.45 & 18 & 13 & 12 & 10 & 12 & 17 \\
CO & 4.46--5.0 & 93 & 62 & 55 & 38 & 52 & 71 \\
\enddata
\tablecomments{The dominant molecules and wavelength ranges for the 7 absorption features in the clear-sky exoVenus spectra. The last 6 columns of the table list the number of transits needed to detect a given feature in each of the 6 exoVenus CO$_2$ cases. Dashes indicate the feature was unable to be detected in less than 100 transit observations.}
\label{tab:T1V}
\end{deluxetable}

\begin{deluxetable}{ c c c c c c c c }
\tabletypesize{\scriptsize}
\tablecaption{Detecting Cloudy ExoVenus Absorption Features}
\tablehead{\colhead{Feature} & \colhead{Wavelengths} &
\multicolumn{6}{c}{Transits Required for Detection} \\
\cline{3-8}
\colhead{} & \colhead{$\mu$m} & \colhead{0.1\%} & \colhead{0.5\%} & \colhead{1\%} & \colhead{25\%} & \colhead{50\%} & \colhead{96\%} }
\startdata
CO$_2$ & 1.16--1.4 & - & - & - & - & - & -   \\
CO$_2$ & 1.4--1.8 & - & - & - & - & - & -   \\
CO$_2$ & 1.83--2.25 & - & - & - & - & - & - \\
CO$_2$ & 2.5--3.08 & - & - & - & 66 & 74 & - \\
SO$_2$ & 3.9--4.11 & - & - & - & - & - & - \\
CO$_2$ & 4.11--4.45 & - & 60 & 53 & 33 & 39 & 65 \\
CO & 4.46--5.0 & - & - & - & - & - & - \\
\enddata
\tablecomments{The dominant molecules and wavelength ranges for the 7 absorption features in the hazy exoVenus spectra. The last 6 columns of the table list the number of transits needed to detect a given feature in each of the 6 exoVenus CO$_2$ cases. Dashes indicate the feature was unable to be detected in less than 100 transit observations.}
\label{tab:T1V_Cloudy}
\end{deluxetable}

\section{Discussion}
\label{discussion}


\subsection{Features to Prioritize in Observations}

If the goal of an observation is to solely detect molecular absorption in an atmosphere, then the 4.3 $\mu$m CO$_2$ feature is likely the best option for both an exoEarth and exoVenus. The feature can be detected in as little as 10 transits in the clear-sky exoEarth and clear-sky exoVenus spectra, and remains detectable with lower atmospheric CO$_2$ and when clouds or haze are present (Figures~\ref{fig:T1E_AllSpec} \& \ref{fig:T1V_AllSpec}). Beyond just confirming the presence of CO$_2$, the CO$_2$ features at 1.7, 2.0, and 2.7 $\mu$m are useful for estimating CO$_2$ abundance. These features only become present in the transmission spectra of atmospheres with high CO$_2$ abundance, whereas the 4.3 $\mu$m feature is visible even in the lowest CO$_2$ cases for both planets (Figures \ref{fig:T1E_AllSpec} and \ref{fig:T1V_AllSpec}). In theory, these features would be useful for estimating whether a planet has a Venus-like or Earth-like amount of atmospheric CO$_2$, and in turn whether the planet may be habitable. The issue however is that these features overlap with adjacent H$_2$O and CH$_4$ absorption bands which could create ambiguity when trying to derive chemical abundances from data with low spectral resolution. Therefore the optimal features that observations should focus on to differ an exoEarth from an exoVenus are those that are unique to either exoVenus or exoEarth spectra and have little overlap with other features. 

The sole exoVenus absorption feature which can not also be found in the exoEarth spectra is the SO$_2$ feature at 4 $\mu$m. Detection of SO$_2$ in atmospheric spectra could be indicative of a dry atmosphere since SO$_2$ is reactive with water vapor, which would inherently rule out the possibility of the planet being Earth-like. Present-day Venus has a dry atmosphere but has reduced amounts of SO$_2$ due to photochemical oxidation caused by UV radiation from the sun. The hypothetical exoVenus in this work orbits TRAPPIST-1, and the reduced UV output from TRAPPIST-1 allows SO$_2$ to have an extended lifetime in the exoVenus atmosphere, which enhances the possibility of the SO$_2$ being detectable. Discovering SO$_2$ on a planet would provide insight into potential volcanism and tectonic activity occurring on the planet, both of which are vital for inferring climate conditions. In the clear-sky exoVenus spectra, the SO$_2$ feature can be detected in as little as 47 transits, and is undetectable in the 50\% and 96\% CO$_2$ cases (Table \ref{tab:T1V}). In the hazy exoVenus spectra however, the SO$_2$ feature is undetectable in all CO$_2$ cases.

\begin{figure*}
\includegraphics[width=\textwidth]{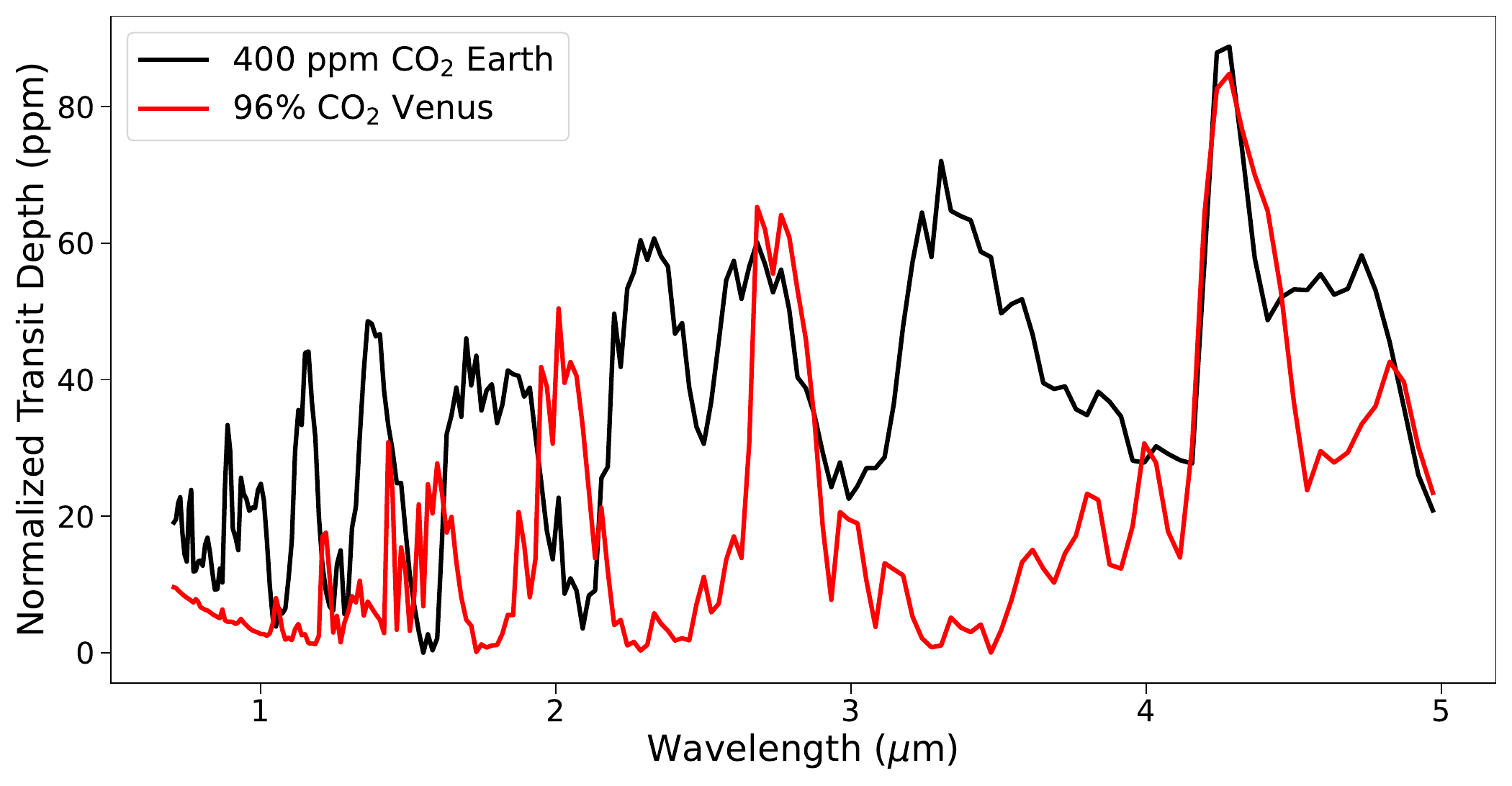}
\caption{The transmission spectra of the 400 ppm CO$_2$ exoEarth and 96\% CO$_2$ exoVenus. There are an abundance of absorption features which are unique to the exoEarth spectrum and could be used to differentiate it from an exoVenus. The exoVenus spectrum yields large CO$_2$ features at 1.7 and 2.0 $\mu$m, however they may be difficult to resolve from adjacent H$_2$O absorption.}
\label{fig:EarthVenusComp}
\end{figure*}

The majority of the CH$_4$ absorption features in the exoEarth spectra do not overlap with CO$_2$ features, and there are no CH$_4$ features present in the exoVenus spectra (Figure~\ref{fig:EarthVenusComp}). Several H$_2$O features are also in the exoEarth spectra but they all have significant overlap with adjacent CO$_2$ features. It has been well documented that detection of CH$_4$ and O$_2$ disequilibrium in an exoplanet atmosphere is a potential biosignatures for identifying habitable worlds \citep[e.g.]{thompson2022case,krissansen2018disequilibrium,schindler2000synthetic,des2002remote,kasting2005methane,lovelock1965physical,hitchcock1967life,arney2017pale}. Although CH$_4$ can contribute to a runaway greenhouse scenario \citep{ramirez2018methane,haqq2008revised,pavlov2000greenhouse}, the likelihood of a detectable amount of CH$_4$ being produced from abiotic processes is far less likely than deriving from biotic processes \citep[e.g.]{guzman2013abiotic,wogan2020abundant}, making it improbable to detect CH$_4$ on an exoVenus. These factors make CH$_4$ the optimal exoEarth absorption feature for discerning an exoEarth from an exoVenus.

The features at 0.9 and 1.1 $\mu$m in the exoEarth spectra are composed of both H$_2$O and CH$_4$ absorption (Table \ref{tab:T1E}). These two features both have slight overlap with CO$_2$ features, but the CO$_2$ are much smaller than the H$_2$O features, even in cases of high CO$_2$ abundance (Figure~\ref{fig:EarthVenusComp}). An additional observational benefit of the exoEarth features at 0.9 and 1.1 $\mu$m is that they are located at wavelengths where the NIRSpec PRISM has the highest spectral resolution Figure~\ref{fig:PandexoExample}. Despite this, the 0.9 $\mu$ feature still requires at least 81 transit observations to be detected. The 1.1 $\mu$m feature proved to be far more detectable, requiring at most 26 transits in all exoEarth cases. 

The feature at 1.8 $\mu$m, which is comprised of roughly equal parts CH$_4$ and H$_2$O absorption, and the CH$_4$ features at 2.3 and 3.5 $\mu$m are all very amenable to detection. All 3 features require at most 9 transits to detect, and have minimal overlap with CO$_2$ features. As a result, these features are optimal for both minimizing observation time and clarifying that a planet is not Venus-like.


\subsection{Observation Time}

To put into perspective the amount of time needed to detect the features in the exoEarth and exoVenus transmission spectra, it is useful to compare the number of transits required for detection to the observation time criteria of JWST observing proposals. JWST proposals are separated into 3 categories based on the length of the observation. Small proposals are limited to 25 hours or less of total observing time, medium proposals range from 25 to 75 hours, and large proposals exceed 75 hours. A single transit observation in our JWST simulations equates to 1.75 hours since we assumed equal observing time in and out of transit. Cycle 2 JWST guest observer (GO) proposals assume an average slew time of 35 minutes, which would make a single transit observation equal to 2.33 hours when including overhead time. While considering slew time a small proposal could include at most 11 transit observations, 32 transits for a medium proposal, and more than 32 transits for a large proposal.

The only exoVenus features which could be detected within a small proposal are the 2.7 $\mu$m CO$_2$ feature in the 25\% CO$_2$ case and the 4.3 $\mu$m CO$_2$ feature in the 1\%, 25\% and 50\% cases. The time allotted by a medium proposal would be sufficient for detecting the 2.0 $\mu$ feature in the 3 highest CO$_2$ cases, and the 2.7 and 4.3 $\mu$m features in all cases. The 1.6 $\mu$m CO$_2$ feature can be detected in the 3 highest CO$_2$ cases in slightly more time than a medium proposal. The 4.0 $\mu$m SO$_2$ feature and 4.8 $\mu$m CO feature would both require a large proposal to be detected in all cases.

A small observing proposal for the exoEarth would be sufficient for detecting the CH$_4$ and H$_2$O features at 1.8, 2.3, 2.8, and 3.5 $\mu$m in all CO$_2$ cases, and the 4.3 $\mu$m CO$_2$ feature in the 4k and 40k ppm CO$_2$ cases. The only feature which would require a large proposal is the CH$_4$ feature at 0.9 $\mu$m, which would need far more transit observations than the minimum for a large proposal.




\section{Conclusions}
\label{conclusions}

In this work we used VPL Climate to model the climate states and atmospheres of 6 hypothetical exoEarths and exoVenuses, all of which are located on the runaway greenhouse boundary in the TRAPPIST-1 system. Both the exoEarths and exoVenuses have varying abundances of atmospheric CO$_2$ that range from 0.4--40k ppm CO$_2$ and 0.1\%--96\% CO$_2$, respectively. The atmospheres from the modelled planets were used as inputs for PSG to simulate their transmission spectra between 0.6--5.3 $\mu$m, including clear-sky and cloudy scenarios for both planets. The transmission spectra were then input into PandExo to simulate observations of the planets using JWST NIRSpec PRISM. Using the simulated JWST data we quantified the detectability of 10 absorption features in the clear-sky and cloudy exoEarth spectra, and 7 features in the clear-sky and hazy exoVenus spectra. From this, we identified features in the spectra of both planets that would be useful for confirming whether a planet is Venus-like or Earth-like. 

The SO$_2$ feature at 4.0 $\mu$m is the only feature unique to the exoVenus spectra, and the best opportunity to confirm a planet is more similar to Venus than Earth. The detectability of the SO$_2$ feature is volatile in clear-sky cases however, as it would require a minimum of 47 transits to detect in low CO$_2$ cases, and becomes undetectable with higher CO$_2$ abundance. When haze was included, the SO$_2$ feature was completely undetectable. The CO$_2$ features at 1.7, 2.0, and 2.7 $\mu$m are indicators for a CO$_2$ dense atmosphere, however they will likely be difficult to discern from the adjacent H$_2$O and CH$_4$ features they overlap with, and are undetectable in hazy atmospheres.

The exoEarth has enhanced atmospheric CH$_4$ abundance because of the SED of its host star. As a result, the CH$_4$ features at 1.15, 1.8, 2.3, and 3.4 $\mu$m are all amenable to detection and are likely the best option in the NIR for ruling out a Venus-like planet. The presence of clouds had a minuscule effect on the detectability of features. Given that the exoVenus spectra only have a single unique absorption feature while the exoEarth spectra have several that are also far more detectable, it will likely be easier to confirm an exoEarth than it will be to confirm an exoVenus. 

JWST is scheduled to observe several VZ planets in cycle 1, and will likely observe more in future cycles. These planets will serve as our first look into the climate conditions of planets with similar insolation flux as Venus, and will be a crucial resource for understanding the causes for Venus being inhospitable today. In addition, the discovery of habitable worlds in the VZ could strengthen the hypothesis that Venus could have had an extended temperate period in its past, and broaden the selection of planets that are targeted in search for signs of life. The importance of exoVenuses to the study of planetary habitability makes the ability to confidently identify Venus-like worlds paramount for maximizing what can be learned from observations of VZ planets with JWST NIRSpec and future facilities like the Extremely Large Telescope (ELT) and the Habitable Worlds Observatory. 


\section*{Acknowledgements}

C.O. and S.R.K. acknowledge support from NASA grant 80NSSC21K1797, funded through the NASA Habitable Worlds Program, and also from NASA grant 80NSSC22M0188, funded through the NASA Discovery Program. P.D. is supported by a National Science Foundation (NSF) Astronomy and Astrophysics Postdoctoral Fellowship under award AST-1903811. A.P.L. was supported by the Virtual Planetary Laboratory Team, which is a member of the NASA Nexus for Exoplanet System Science, and funded via NASA Astrobiology Program Grant 80NSSC18K0829.  The results reported herein benefited from collaborations and/or information exchange within NASA's Nexus for Exoplanet System Science (NExSS) research coordination network sponsored by NASA's Science Mission Directorate. This work made use of the advanced computational, storage, and networking infrastructure provided by the Hyak supercomputer system at the University of Washington. The authors thank Dr. Tara Fetherolf and Dr. Kimberly Bott for their feedback which helped improve the quality of this work.


\software{Planetary Spectrum Generator \citep{villanueva2018planetary}, PandExo: ETC for Exoplanets \citep{batalha2017pandexo} }


\bibliographystyle{aasjournal}
\bibliography{references}


\end{document}